\documentclass[aps,prb,groupedaddress]{revtex4}

\usepackage{graphicx}
\usepackage{dcolumn}
\usepackage{epsfig}
\usepackage{bm}
\usepackage{amsfonts}
\usepackage{latexsym}

\newcommand{\sbfdelta}{\mbox{{\scriptsize\boldmath $\delta$}}}
\newcommand{\be}{\begin{equation}}
\newcommand{\ee}{\end{equation}}
\newcommand{\bea}{\begin{eqnarray}}
\newcommand{\eea}{\end{eqnarray}}

\newcommand{\half}{\frac{1}{2}}
\newcommand{\figwidths}{0.325\columnwidth}
\newcommand{\figwidth}{0.55\columnwidth}

\begin{document}


\title{The Mixed Spin $S=(\half,1)$ XXZ Ferrimagnet at Zero Temperature}



\author{Weihong Zheng}
\email[]{w.zheng@unsw.edu.au}
\homepage[]{http://www.phys.unsw.edu.au/~zwh}
\affiliation{School of Physics,
The University of New South Wales,
Sydney, NSW 2052, Australia.}

\author{J. Oitmaa}
\email[]{j.oitmaa@unsw.edu.au}
\affiliation{School of Physics,
The University of New South Wales,
Sydney, NSW 2052, Australia.}

%

\date{\today}

\begin{abstract}
Linked cluster series expansions about the Ising limit are used to study ground
state preperties, viz. ground state energy, magnetization and excitation spectra,
for mixed spin $S=(\half,1)$ quantum ferrimagnets on simple bipartite
lattices in 1, 2, and 3-dimensions. Results are compared to second-order spin
wave theory and, in general, excellent agreement is obtained.
\end{abstract}


\maketitle

\section{\label{sec:intro}INTRODUCTION}

Ferrimagnets are materials where ions on different sublattices have opposing
magnetic moments which do not cancel in the ordered phase\cite{wol61}.
This can arise either through unequal numbers of ions on the sublattices
or the ions having different spin quantum numbers. There is growing interest in
such systems, both from fundamental physics and through their technological
potential. Arguably the simplest such structures are bipartite lattices
(A,B) with $S_A\neq S_B$, with nearest neighbour antiferromagnetic exchange
coupling. The experimental discovery of ferrimagnetism in bimetallic
chains\cite{kon90} has led to many studies of mixed-spin chains\cite{pat97,yam98,wu99}.
The rare-earth nickelates R$_2$BaNiO$_5$\cite{gar95}, which can be modelled by a
2-dimensional net of coupled chains\cite{tak00,alv02}, are an example of a mixed-spin
system in 2-dimensions. Another example is an Fe-Ni cyanide bridged network\cite{par02},
which provides a realization of the square lattice model we consider in this paper.

We consider, in this paper, the cases of a bipartite chain, sqaure lattice, and
simple-cubic lattice where sublattice $A$ is occupied by $S=\half$ spins and
sublattice $B$ by $S=1$ spins, with nearest neighbour antiferromagnetic coupling
between the sublattices. This is the extreme quantum limit, where quantum fluctuations
will be most significant. The Hamiltonian is taken as
\be
H = J \sum_{\langle ij\rangle} [ s_i^z S_j^z + \frac{x}{2}
( s_i^{+} S_j^{-} + s_i^{-} S_j^{+} ) ]
\ee
where the parameter $0<x<1$ represents exchange anisotropy, the coupling
$J>0$ and it can be set to be 1.

Previous work on this system includes spin-wave theory\cite{iva00}, and a variety of
systematic numerical methods\cite{pat97,yam98,wu99}, with
most emphasis on the mixed-spin chain. Other work has included
frustrating further neighbour interactions for the chain\cite{iva98} and
square lattice\cite{iva02}, but we do not consider such cases here.

Our approach is to use high-order linked-cluster expansions\cite{gel00},
where quantities are expanded about the Ising limit as power series in $x$,
a technique we have employed in previous studies of quantum antiferromagnets\cite{zhe91}.
This allows the whole region $0<x<1$ to be studied, as well as extrapolation to
the isotropic Hamiltonian $x=1$. In particular we compute series for the
ground state energy, the magnetization on each sublattice, and
the dispersion relations for magnon excitations.

Since we will compare the series predictions with spin-wave theory,
we previde here the relevant spin-wave results at lowest order, for
the general case with spins $S_1$ and $S_2$. Further details,
including the second-order results, are given in the Appendix.
The ground state energy and magnetizations are
\bea
E_0/NJ &=& - z S_1 S_2/2 + \frac{z}{2N} \sum_{\bf k} [ \sqrt{(S_1+S_2)^2-4 S_1 S_2 x^2
\gamma_{\bf k}^2 } - (S_1 + S_2) ] \label{eq2} \\
M_{1,2} &=& S_{1,2} - {1\over N} \sum_{\bf k} [ {S_1+S_2 \over \sqrt{ (S_1 + S_2)^2
 - 4 S_1 S_2 x^2 \gamma_{\bf k}^2}} -1 ]  \label{eq3}
\eea
where the sum is over $N/2$ k-values in the reduced Brillouin zone, $z$ is the
lattice coordination number, and $\gamma_{\bf k}$ is the usual factor
$\gamma_{\bf k}=\frac{1}{z}\sum_{\delta} e^{i {\bf k}\cdot \sbfdelta} $.
The first term in (\ref{eq2}), (\ref{eq3}) is the classical result, and
the second term is the lowest correction due to quantum fluctuations.

The two branches of magnon energies are given by
\be
\omega_k^{\pm} = \frac{1}{2}zJ [ \sqrt{ (S_1 + S_2)^2 - 4 S_1 S_2 x^2 \gamma_{\bf k}^2 }
~\pm (S_2-S_1) ]
\ee
The most interesting feature of these is that, at $x=1$, one branch is gapped,
while the other becomes gapless but with quadratic dispersion
\be
\begin{array}{l}
\omega_{\bf k}^+ \simeq z J (S_2-S_1) \\
\omega_{\bf k}^- \simeq {\rm const.} ~k^2
\end{array} {\Big \}} \quad {\bf k} \to 0
\ee
This result is well known, and is a consequence of $S_1\neq S_2$.
Another feature, which we have not seen commented on previously, is that
the various quantities are analytic at $x=1$, unlike the case of
the usual antiferromagnet, where there are square root singularities.
This make the extrapolation of our series to $x=1$ much more precise.

The outline of the paper is as follows. In Section II we
give (brief) details of our method and present some of the
raw data. In Section III we present an analysis, mainly in graphical
format, for 1, 2 and 3-dimensions. The spin wave results are shown
for comparison. We find that many of our results agree extremely
well with second-order spin-wave theory, and are indistinguishable
on the scale of the fingures. In Section IV we present
some conclusions.

\section{\label{sec2}Derivation of Series Expansions}

The series are obtained, as described in ref.[\onlinecite{gel00}],
by evaluating the ``proper" or ``cumulant" contribution for each
of a set of connected clusters, and then summing these, with their
respective embedding factors, to obtain the series for the bulk
system. In this way the ground state energy per site, for example,
is written in the form
\be
E_0/NJ = \sum_{n=0}^{\infty} e_n x^n
\ee
where the purely numerical coefficients $e_n$ are computed exactly
to some maximum order (in this work, to order 22, 14, 12 for
$d=1$, 2, 3 respectively). The coefficients are given
in Table I, for the case $S_1=\frac{1}{2}$, $S_2=1$, which is the only
case we consider. Table II gives the corresponding coefficients for the
magnetization series. We note that, apart from the constant term,
these series are identical and hence only $M_1$ is given.

Series for the excitation spectra are obtained via the linked
cluster method of Gelfand (see [\onlinecite{gel00}]), in which we write
\be
\omega ({\bf k}) = \sum_{\bf R} t({\bf R}) e^{i {\bf k}\cdot {\bf R} }
\ee
where the $t({\bf R})$ are a set of "transition weights" obtained from
an effective Hamiltonian. The series for two branches of
excitation spectra has been computed
upto order 22, 10, 8 for
$d=1$, 2, 3 respectively, which involve a list of 22, 185928, 59804
clusters, respectively.
In Table II we provide our results for these,
for the square lattice. In Table I, we also list the series
for excitation spectra at ${\bf k}=0$ for all three lattices.
We are happy to provide other
results on request.

\section{\label{sec3}Results}
The series can be evaluated by Pad\'e approximants or integrated
differential approximants\cite{gut}, either for a particular $x$ or extrapolated to
$x=1$. We display our results in graphical form.

\subsection{Ground state energy}
\begin{figure}[t]
  { \centering
    \includegraphics[width=\figwidths]{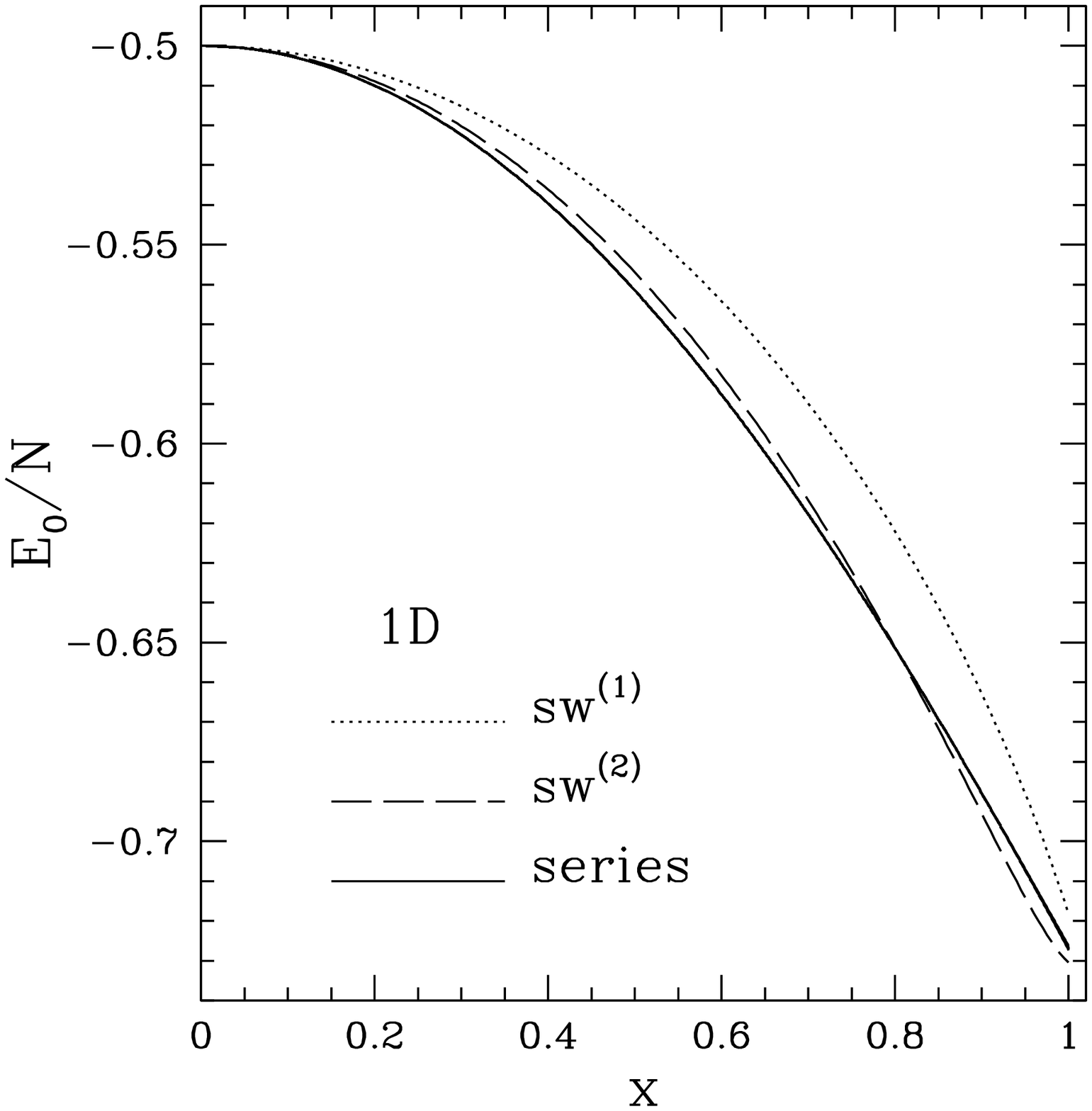} \hspace{-0.05cm}
    \includegraphics[width=\figwidths]{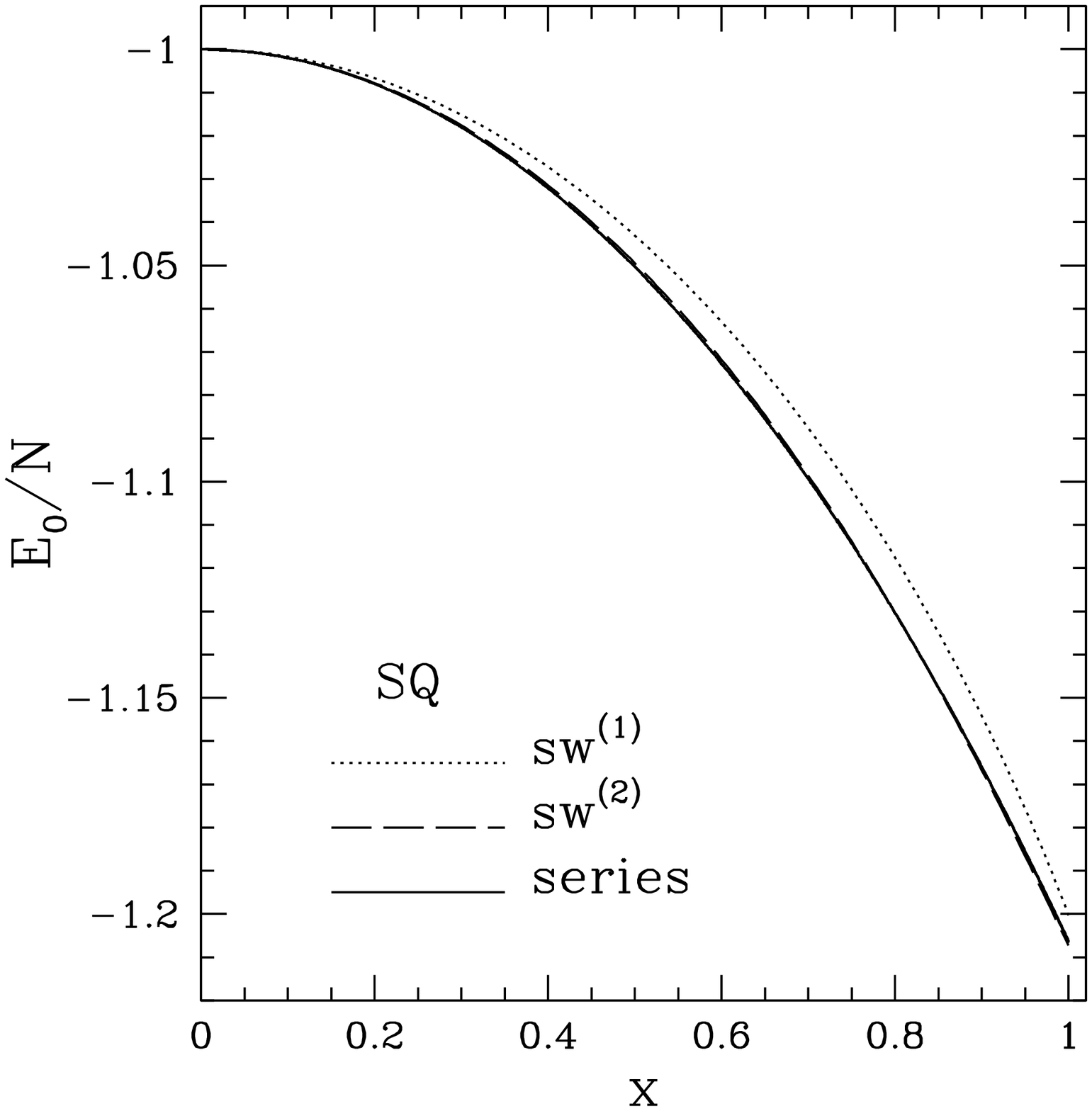} \hspace{-0.05cm}
    \includegraphics[width=\figwidths]{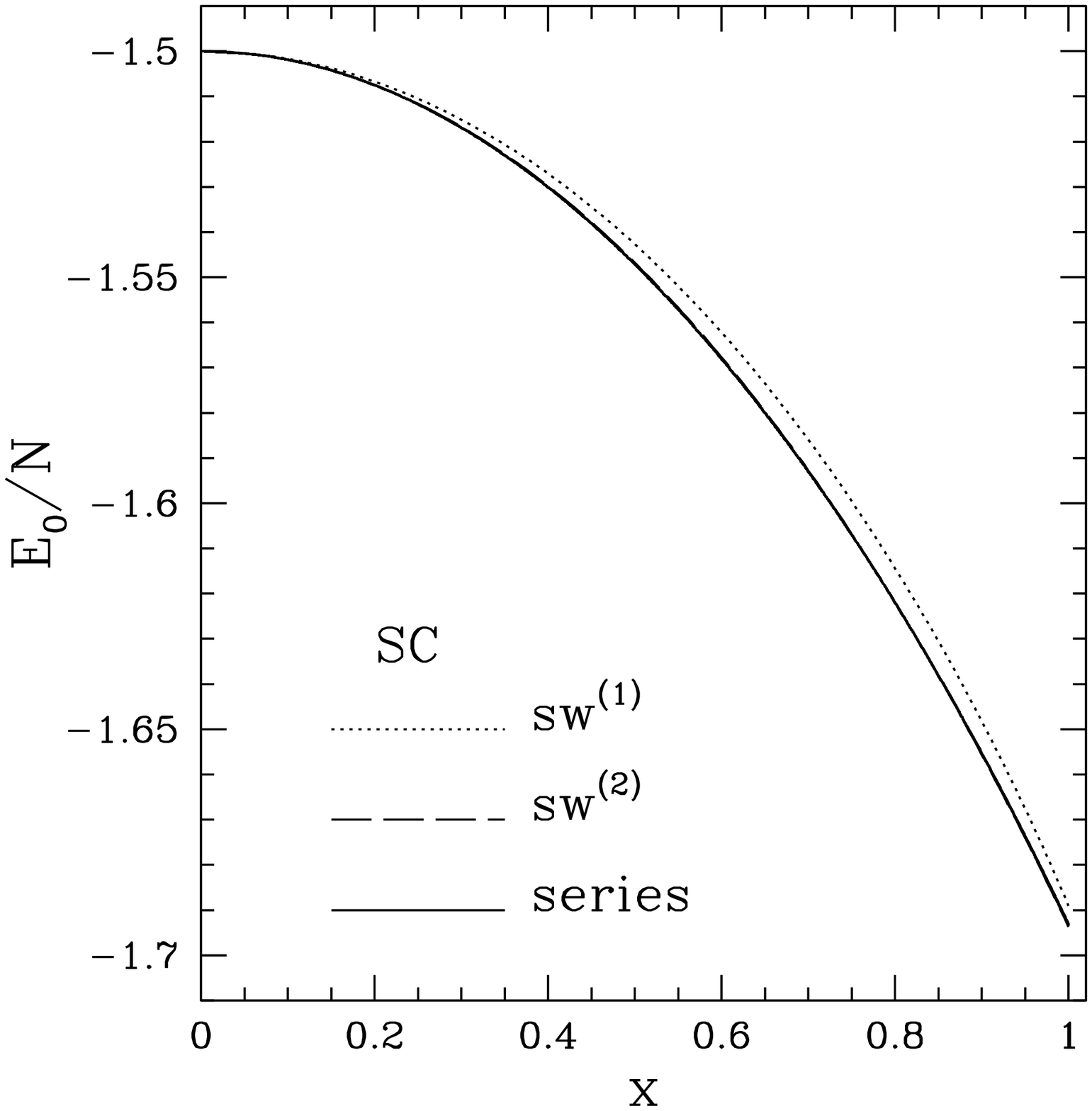}
    \caption{Ground state energy per site for the $S=(\half,1)$
    system on linear chain (a), square lattice (b) and simple cubic lattice (c).
    The dotted and dashed lines, in each figure are the results of first-order
    and second-order spin-wave theory.
    \label{fig_e0}}
  }
\end{figure}

Figure \ref{fig_e0} shows the ground state energy per site, as a function
of exchange anisotropy $x$, for the three lattices.

We note from this figure the excellent agreement between our series results
and, particularly, second-order spin-wave theory. The agreement is not perfect
for the chain but for $d=2,3$ the curves are indistiguishable on the scale
of the figures. The estimated numerical error in the series is less than
the width of the curve and so these results are essentially numerically exact.

\begin{figure}[t]
  { \centering
    \includegraphics[width=\figwidths]{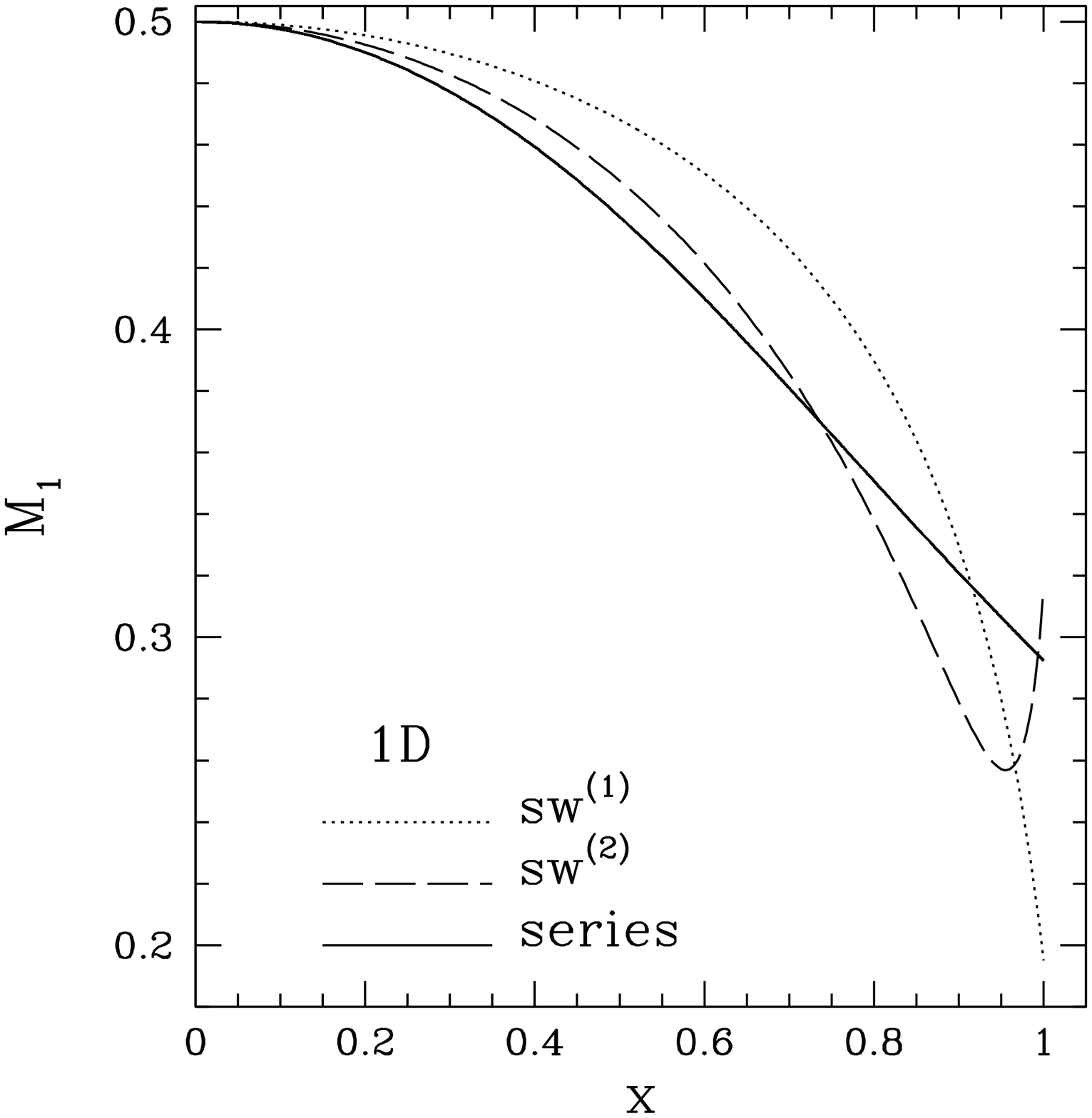} \hspace{-0.05cm}
    \includegraphics[width=\figwidths]{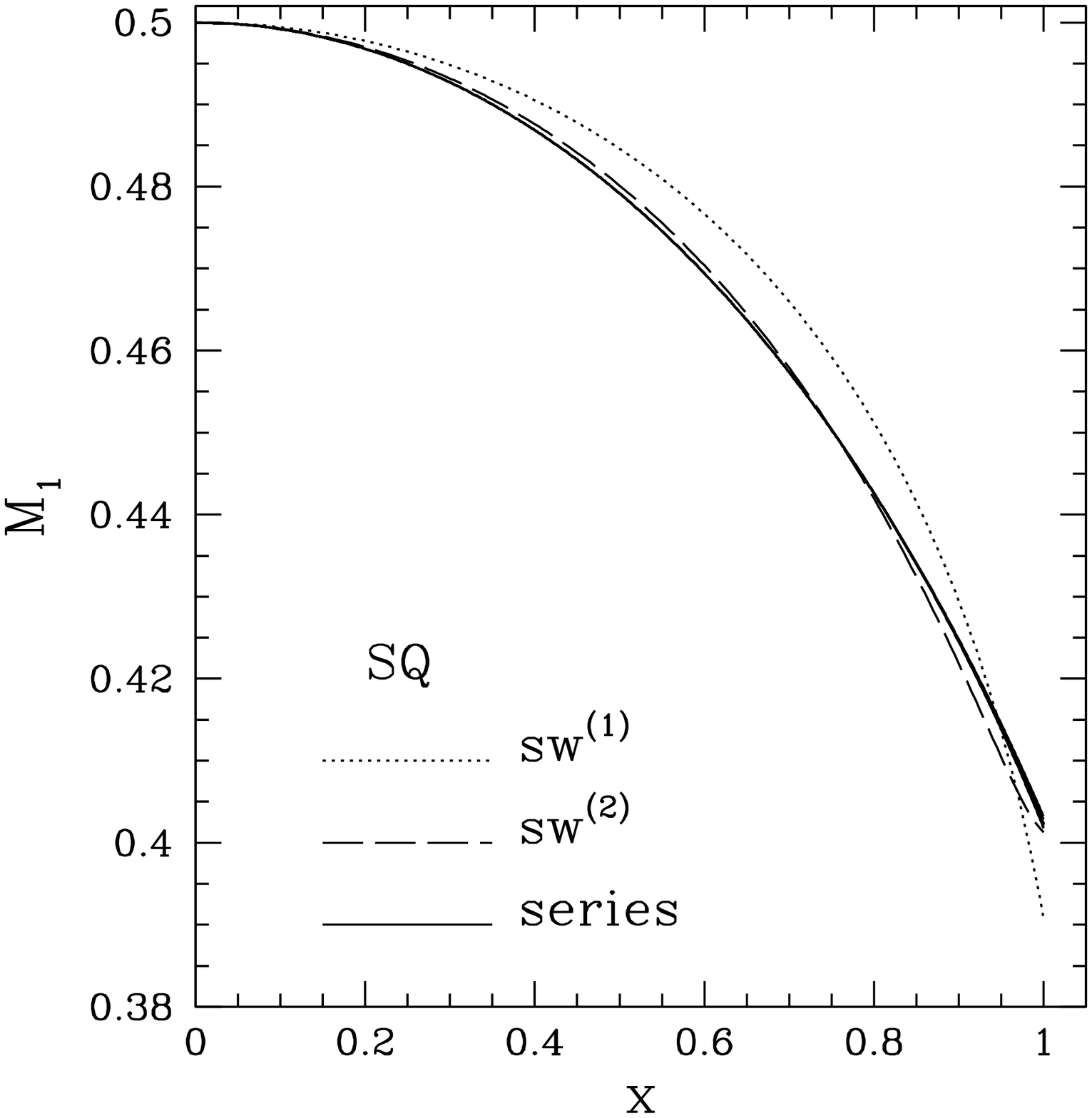} \hspace{-0.05cm}
    \includegraphics[width=\figwidths]{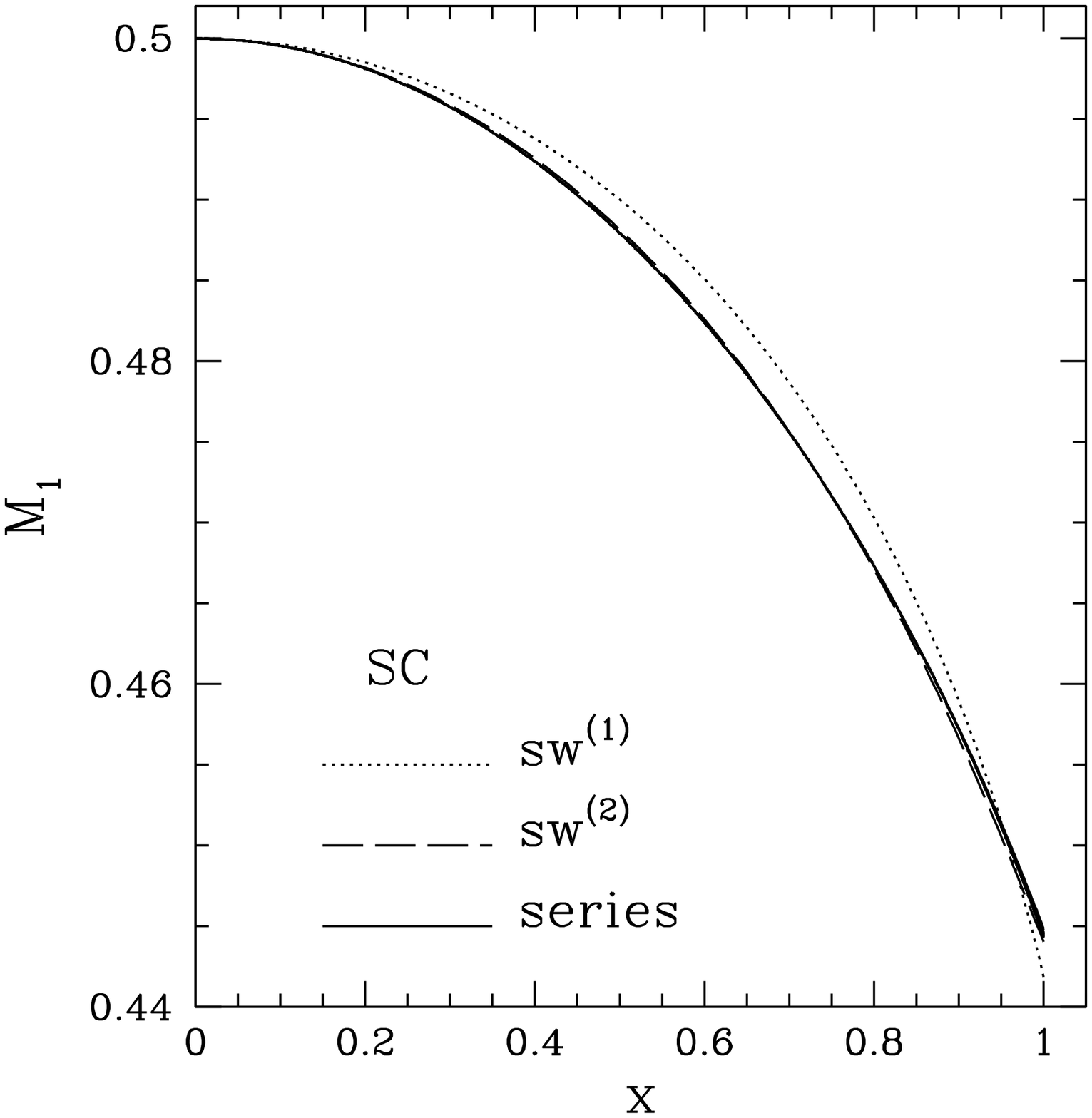}
    \caption{Magnetization per site on the $S=\half$ sublattice for
    the linear chain (a), square lattice (b) and simple cubic lattice (c).
    The dotted and dashed lines, in each figure are the results of first-order
    and second-order spin-wave theory.
    \label{fig_M}}
  }
\end{figure} 

Figure \ref{fig_M} shows the magnetization as a function of anisotropy
$x$, again for the three lattices.
Let us start from the $d=3$ case, where spin wave theory would be expected to
be most accurate. We see that this is indeed the case, and second order spin-wave theory
is barely distinguishable from our series results. At the isotropic point
$x=1$ quantum fluctuations reduce the moment from 0.5 to 0.445, an 11\% reduction
(actually our numerical result is 0.44450(3)).
For the square lattice quantum fluctuations reduce the sublattice moment
by about 20\% at $x=1$.
The series results are close to, but distinguishable from,
second-order spin-wave theory.
For the linear chain the differences are greater and spin-wave theory does not perform well,
especially approaching $x=1$. The series results are still very precise,
and give a sublattice moment of  $\sim 0.29$  at the isotropic
point, i.e. a 40\% reduce due to quantum fluctuations
(the numerical value is $0.292487(6)$).
We note here that Tian\cite{tia97} has show rigorously that this
system will have long range order at $T=0$, even in 1-dimension.
First-order spin wave theory overestimates the effect of quantum fluctuations
near $x=1$, while the second-order curve shows a dip and then a sharp rise
near $x=1$. This is clearly a spurious effect.

\subsection{Exitated states}

\begin{figure}[t]
  { \centering
    \includegraphics[width=\figwidth]{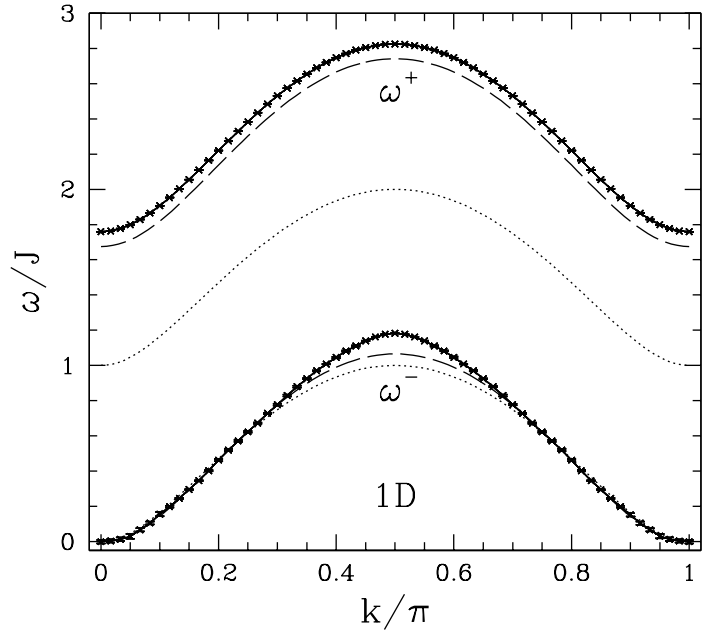}
    \caption{Excitation energies versus $k$ for the $s=(\half,1)$ system on the
    linear chain.
    The dotted and dashed lines give the corresponding first-order
    and second-order spin-wave theory.
    \label{fig_mk_1d}}
  }
\end{figure}

Finally we turn to the excitation spectra. These are only shown for
the isotropic case, $x=1$. Figure \ref{fig_mk_1d} shows
the two excitation branches for the chain versus $k$,
again together with the spin-wave results. The $k^2$ dispersion is evident in
both spin wave and series curves. Both bands have an approximante
cosine shape. For the lower branch spin wave theory is quite accurate, except
near $k={\pi\over 2}$, but for the upper branch first-order spin-wave
theory gives a large underestimate. Second-order theory is better, but still
under estimates the excitation energy over the entire $k$ range.

\begin{figure}[t]
  { \centering
    \includegraphics[width=\figwidth]{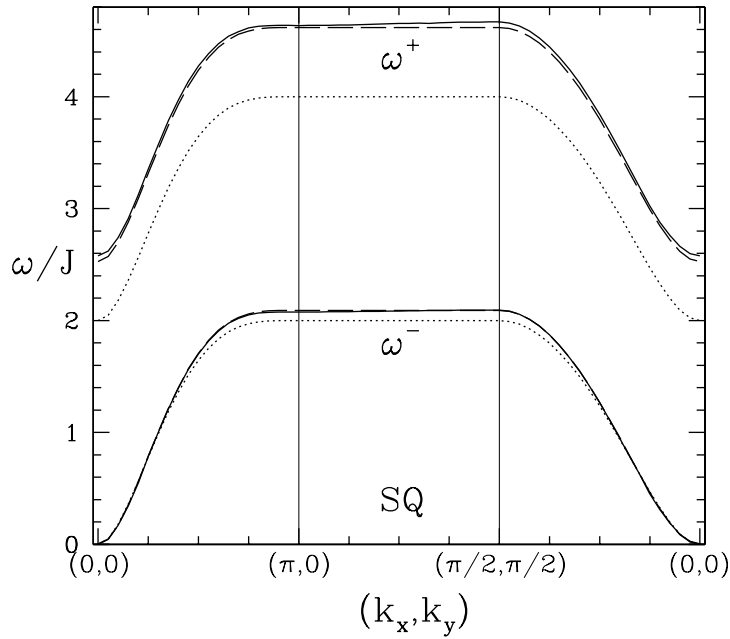}
    \caption{Magnon dispersion curves for the $s=(\half,1)$ system on
    the square lattice, for special lines in the Brillouin zone.
    The dotted and dashed lines give the  first-order
    and second-order spin-wave results.
    \label{fig_mk_sq}}
  }
\end{figure}

Figure \ref{fig_mk_sq} shows the dispersion curves for magnon excitation
for the square lattice, along 3 lines in the Brillouin zone. In this case the
series and second-order spin-wave results agree very well at all points,
while the first-order spin-wave results are too low. Note the almost flat
spectrum along the zone boundary from $(\pi,0)$ to $(\pi/2,\pi/2)$.
In first and second order spin-wave theory this is completely flat, as
the ${\bf k}$-dependence arises from $\gamma_{\bf k} =\frac{1}{2} (\cos k_x +
\cos k_y )$. The series result shows a slight rise from
$(\pi,0)$ to $(\pi/2,\pi/2)$.

\begin{figure}[t]
  { \centering
    \includegraphics[width=\figwidth]{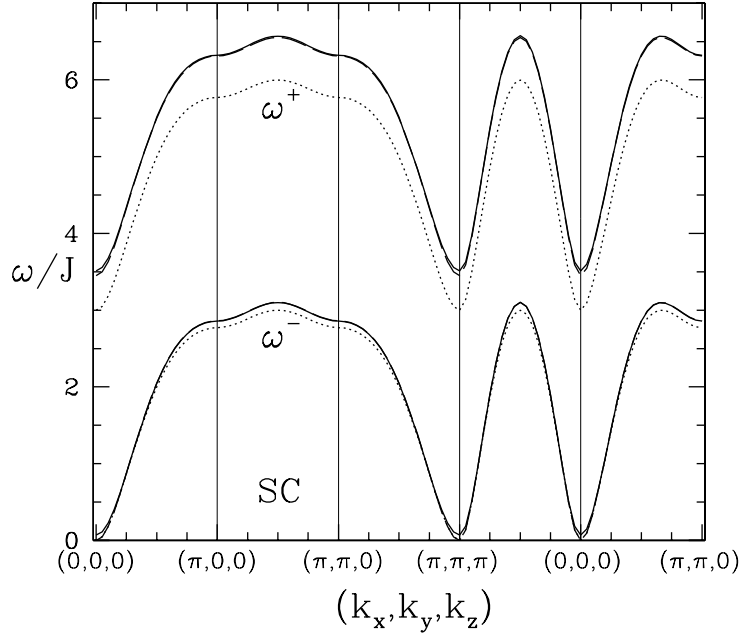}
    \caption{Magnon dispersion curves for the $s=(\half,1)$ system on
    the simple cubic lattice, for special lines in the Brillouin zone.
    The dotted and dashed lines give the  first-order
    and second-order spin-wave results.
    \label{fig_mk_sc}}
  }
\end{figure}

The corresponding results for the simple cubic lattice are shown
in Figure \ref{fig_mk_sc}, again for special lines in the zone. Agreement between
series and spin-wave results is excellent, with the second order spin-wave
results virtually indistiguishable from the series.

\section{\label{sec4}Conclusions}


We have used linked cluster series expansions about the Ising limit to investigate
ground state properties of simple mixed spin $S=(\half, 1)$ quantum
antiferromagnets, on linear chain, square and simple cubic lattices.
The series are very regular and easily analyzable, and we believe
the results are essentially numerically exact. We have calculated the ground
state energy, sublattice magnetizations, and magnon excitation spectra,
and compared our results with first and second order spin wave
 theory. As is known from previous work on Heisenberg antiferromagnets,
 second-order spin-wave theory is surprisingly accurate, even in 1-dimension.
 In the present case the agreement is even better, as demonstrated by our
 results above. Presumably this reflects the fact that the
 ground state has a finite spin.

Our calculated excitation spectra show a quadratic dispersion at ${\bf k}=0$, and we
have also computed the curvature $\alpha$, defined by $\omega^-_k = \alpha k^2$.
In Table II we summarize all our numerical results at $x=1$, and
present a comparison with other methods.

 This work demonstrates that the series method is very powerful in studying
 mixed spin quantum ferrimagnets, and encourages its use in more complex mixed-spin
 models, including, for example, frustrated systems. We are pursuing
 a number of calculations along these lines.

\appendix
\subsection*{Appendix}

We provide, for completeness, a summary of first and second order
spin wave theory for this system. Equivalent treatments have been given in
Ref. \onlinecite{iva00}, and elsewhere.

The Hamiltonian for the anisotropic system is (setting $J=1$)
\begin{equation}
H= \sum_{\langle lm \rangle} \left[ s^z_l S^z_m +x(s^x_lS^x_m+
s^y_l S^y_m)\right] + h_1\sum_l s^z_l + h_2 \sum_m S^z_m , \label{h0}
\end{equation}
where we have divided the lattice sites into even and odd sublattices,
denoted
by $l$ and $m$ respectively, and  introduced the magnetic fields $h_1$ and $h_2$.

We firstly introduce boson operators $a_l$ and $b_m$ via the Dyson-Maleev
transformation on the two sublattices:
\begin{eqnarray}
{\rm l-sublattice:}&&  ~~s^z_l = S_1\!-\!a^{{\dag}}_l a_l, ~s^+_l =
(2S_1)^{1/2} a_l \!-\! (2S_1)^{-1/2} a^{\dag}_la_la_l, ~s^-_l=(2S_1)^{1/2}
a^{\dag}_l, \nonumber \\ && \\
{\rm m-sublattice:}&& ~~S^z_m =b^{{\dag}}_m b_m\!-\!S_2,~S^+_m =
(2S_2)^{1/2} b_m^{\dag}\! -\! (2S_2)^{-1/2} b^{\dag}_m b_m^{\dag} b_m,~
S^-_m=(2S_2)^{1/2} b_m.  \nonumber
\end{eqnarray}
Note that this transformation is not Hermitian. In terms of the
boson operators, the Hamiltonian can be expressed as:
\begin{eqnarray}
H &=& -N(S_1 S_2 z - h_1 S_1 + h_2 S_2)/2  \nonumber \\
&& + (zS_2-h_1) \sum_l a^{\dag}_l a_l + (zS_1+h_2)\sum_m b^{\dag}_m b_m  +
x\sqrt{S_1 S_2} \sum_{\langle lm \rangle} ( a_l b_m + a^{\dag}_l b_m^{\dag}) \nonumber \\
&& -\sum_{\langle lm \rangle} a_l^{\dag} a_l b_m^{\dag} b_m -{x\over 2}
\sqrt{S_1 S_2} \sum_{\langle lm \rangle} ( a_l^{\dag} a_l a_l b_m/S_1 + a_l^{\dag}
b_m^{\dag} b_m^{\dag} b_m/S_2 ) ~.
\end{eqnarray}
Then, as in Ref. \onlinecite{zhe91}, we introduce the Bloch-type boson operators $a_k$,
$b_k$ by a Fourier transformation:
\begin{equation}
a_k = \left( {2\over N} \right)^{1/2} \sum_{l} e^{ik\cdot l} a_l, \quad
b_k = \left( {2\over N} \right)^{1/2} \sum_{m} e^{-ik\cdot m} b_m,
\end{equation}
where $N$ is the total number of lattice sites. The quadratic part of $H$
can be diagonalized by a Bogoliubov transformation:
\begin{eqnarray}
a_k &=& \alpha_k\cosh \theta_k-\beta^{\dag}_k\sinh \theta_k,
\nonumber  \\    & & \\
b_k & =& -\alpha_k^{\dag}\sinh\theta_k+\beta_k\cosh\theta_k, \nonumber
\end{eqnarray}
where $\tanh 2\theta_k=x\gamma_k/D$, and
$ D=( z S_1 + z S_2 - h_1 + h_2 )/( 2 z \sqrt{S_1 S_2})$.

With this, one can get the ground state energy up
to second order in $1/S$ expansion:
\begin{eqnarray}
E_h/N &=& (-z S_1 S_2 + h_1 S_1 + h_2 S_2)/2 + z \sqrt{S_1 S_2} D C^h_1/2 \nonumber \\
&& - {z\over 8} \Big \{ [ C^h_{-1} + D_0 D (C^h_1 - C^h_{-1} )]^2
+ x^{-2} D^2 (1-x^2 D_0^2 ) (C^h_{-1} - C^h_1)^2 \Big \} ,
 \label{eh}
\end{eqnarray}
where $D_0 = D(h_1=h_2=0) = (S_1 + S_2)/(2 \sqrt{S_1 S_2})$,
$C^h_n = {2\over N} \sum_{\bf k} [ (1-x^2 \gamma_{\bf k}^2 D^{-2} )^{n/2} -1] $.

Setting the external magnetic field to zero (i.e. $h_1=h_2=0$), one can derive
from Eq.(\ref{eh}) the ground-state energy $E_0$:
\be
E_0/N = -z S_1 S_2/2 + z (S_1 + S_2) C_1/4
- {z\over 8}
\Big \{ [C_{-1} + D_0^2 (C_1 - C_{-1})]^2 + x^{-2} D_0^2 (1-x^2 D_0^2) (C_{-1} - C_1)^2
\Big \}
\ee
where
\begin{equation}
C_n= C^{h=0}_n = {2\over N} \sum_{\bf k} \left[ (1-x^2\gamma^2_{\bf k} D_0^{-2} )^{n/2}-1\right].\label{cn}
\end{equation}

Differentiating Eq.(\ref{eh}) with respect to $h_1$,  one
finds the  magnetization $M_1$:
\be
M_1= {2\over N} {\partial E_h \over \partial h_1}{\Big\vert}_{h_1=h_2=0}=
S_1-{C_{-1}\over 2} + {C_{-3} - C_{-1} \over 2x^2 (S_1 + S_2)}
[ D_0^2 (1-x^2) (C_1 - C_{-1}) + x^2 (D_0^2-1) C_{-1} ] ,
\ee
Similarity, one can get $M_2$, which is different from $M_1$ by
an constant $S_2-S_1$, as expected.

One can also get the two branches of magnon dispersion without the
external magnetic field, again up
to second order in a $1/S$ expansion:
\bea
\omega^{\pm}/J &=& \pm z (S_2 - S_1)/2 + z (S_1 + S_2) Q_1/2
\pm { D_0 z (S_2-S_1) \over 4 \sqrt{S_1 S_2}} (C_{-1} - C_1 ) \nonumber \\
&& - {z\over 2} {\Big [} (x^{-2} - 1) D_0^2 (Q_{-1} - Q_1 ) (C_{-1} - C_1 ) + Q_1 C_1 D_0^2
+ Q_{-1} C_{-1} (1-D_0^2) {\Big ]}
\eea
where
$Q_n = (1-x^2 \gamma_{\bf k}^2 D_0^{-2} )^{n/2}$

For $x=1$, we recover the second-order
spin-wave results of Ivanov\cite{iva98}.
It is interesting to study the asymptotic behaviour for the quantities $C_n$
near $x=1$. For the case $S_1=S_2$, the leading order correction term
to $C_n$
near $x=1$ is\cite{zhe91} ${\rm const~ }\times (1-x^2)^{1/2}$,
while for case $S_1\neq S_2$, it is wasy to show that the
asymptotic behaviour for the quantities $C_n$
near $x=1$ is:
\be
C_n (x) = C_n (1) + \frac{n}{2} [C_{n-2}(1) - C_n(1) ] (1-x^2) + \cdots
\ee

For linear chain (1D), square lattice (SQ) and simple cubic lattice (SC),
$C_n$ at $x=1$ for $(S_1,S_2)=(1/2,1)$ are
\bea
&{\rm 1D~} & C_{1} = -0.29097039335, ~C_{-1} = 0.60977301072, ~C_{-3} = 5.3812664598 \\
&{\rm SQ~} & C_{1} = -0.13351109261, ~C_{-1} = 0.21847549012, ~C_{-3} = 1.3753265565 \\
&{\rm SC~} & C_{1} = -0.08401123644, ~C_{-1} = 0.11629596999, ~C_{-3} = 0.5647562506 \\
\eea

\begin{acknowledgments}
This work  forms part of
a research project supported by a grant
from the Australian Research Council.
The computations were performed on an AlphaServer SC
 computer. We are grateful for the computing resources provided
 by the Australian Partnership for Advanced Computing (APAC)
National Facility.
\end{acknowledgments}

\newpage
\bibliography{basename of .bib file}


\begin{table*}
\squeezetable
\caption{Series coefficients for  the ground state energy per site $E_0/NJ$
Magnetization $M_1$, and two branches of magnon $\omega^\pm/J$ at ${\bf k}=0$
for the linear chain, the
square lattice and the simple cubic lattice.
Nonzero coefficients $x^r$
up to order $r=22$ for 1d, order 14 for square lattice, and order 12 for simple cubic
lattice  are listed.}\label{tab_ser}
\begin{ruledtabular}
\begin{tabular}{|rllll|} 
\multicolumn{1}{|c}{$r$} & \multicolumn{1}{c}{$E_0/NJ$} & \multicolumn{1}{c}{$M_1$}
 & \multicolumn{1}{c}{$\omega^-({\bf k}=0)/J$} & \multicolumn{1}{c|}{$\omega^+({\bf k}=0)/J$}  \\
\hline
\multicolumn{5}{|c|}{linear chain}\\
  0 & -5.00000000$\times 10^{-1}$ & ~5.00000000$\times 10^{-1}$ & ~1.00000000  & ~2.00000000  \\
  2 & -2.50000000$\times 10^{-1}$ & -2.50000000$\times 10^{-1}$ & -2.00000000  & -1.00000000  \\
  4 & ~1.04166667$\times 10^{-2}$ & -5.90277778$\times 10^{-2}$ & ~1.12500000  & ~2.31250000  \\
  6 & ~2.56799769$\times 10^{-2}$ & ~2.01786748$\times 10^{-1}$ & ~3.45052083  & -5.81163194  \\
  8 & -1.40457216$\times 10^{-2}$ & -7.01659922$\times 10^{-2}$ & -3.34128599  & ~1.92529744$\times 10^{+1}$ \\
 10 & -6.52895221$\times 10^{-3}$ & -1.51931908$\times 10^{-1}$ & -1.24692841$\times 10^{+2}$ & -7.43345195$\times 10^{+1}$ \\
 12 & ~1.20231578$\times 10^{-2}$ & ~1.67134766$\times 10^{-1}$ & ~9.95763258$\times 10^{+2}$ & ~3.08779448$\times 10^{+2}$ \\
 14 & -9.68145478$\times 10^{-4}$ & ~6.73415017$\times 10^{-2}$ & -3.98363592$\times 10^{+3}$ & -1.34245308$\times 10^{+3}$ \\
 16 & -1.00328363$\times 10^{-2}$ & -2.45737502$\times 10^{-1}$ & ~7.07316949$\times 10^{+3}$ & ~6.03441420$\times 10^{+3}$ \\
 18 & ~6.99312641$\times 10^{-3}$ & ~8.82808138$\times 10^{-2}$ & ~6.98417467$\times 10^{+3}$ & -2.78235446$\times 10^{+4}$ \\
 20 & ~5.40374090$\times 10^{-3}$ & ~2.41826850$\times 10^{-1}$ & ~5.10317554$\times 10^{+4}$ & ~1.30865902$\times 10^{+5}$ \\
 22 & -1.04177497$\times 10^{-2}$ & -2.78535289$\times 10^{-1}$ & -1.73937095$\times 10^{+6}$ & -6.25428515$\times 10^{+5}$ \\
\hline
\multicolumn{5}{|c|}{square lattice}\\
  0 & -1.00000000  & ~5.00000000$\times 10^{-1}$ & ~2.00000000  & ~4.00000000  \\
  2 & -2.00000000$\times 10^{-1}$ & -8.00000000$\times 10^{-2}$ & -1.90000000  & -1.40000000  \\
  4 & -5.22222222$\times 10^{-3}$ & -1.25320988$\times 10^{-2}$ & -1.08660714$\times 10^{-2}$ & ~2.88214286$\times 10^{-2}$ \\
  6 & -7.77432120$\times 10^{-4}$ & -2.94020395$\times 10^{-3}$ & -6.48931742$\times 10^{-2}$ & -3.05217039$\times 10^{-2}$ \\
  8 & -3.41477688$\times 10^{-4}$ & -1.37128294$\times 10^{-3}$ & -9.96489680$\times 10^{-3}$ & -1.95162662$\times 10^{-2}$ \\
 10 & -9.77862982$\times 10^{-5}$ & -5.45189634$\times 10^{-4}$ & -6.07211572$\times 10^{-3}$ & ~5.18453462$\times 10^{-4}$ \\
 12 & -4.20558132$\times 10^{-5}$ & -2.71359073$\times 10^{-4}$ \\
 14 & -1.75132661$\times 10^{-5}$ & -1.33427314$\times 10^{-4}$ \\
\hline
\multicolumn{5}{|c|}{simple cubic lattice}\\
  0 & -1.50000000  & ~5.00000000$\times 10^{-1}$ & ~3.00000000  & ~6.00000000  \\
  2 & -1.87500000$\times 10^{-1}$ & -4.68750000$\times 10^{-2}$ & -2.46428571  & -2.03571429  \\
  4 & -3.89229911$\times 10^{-3}$ & -4.79023836$\times 10^{-3}$ & -2.21734002$\times 10^{-1}$ & -2.12222866$\times 10^{-1}$ \\
  6 & -1.24575187$\times 10^{-3}$ & -1.91047285$\times 10^{-3}$ & -1.50472245$\times 10^{-1}$ & -1.23920805$\times 10^{-1}$ \\
  8 & -4.21416977$\times 10^{-4}$ & -9.04898794$\times 10^{-4}$ & -5.75189462$\times 10^{-2}$ & -5.89627745$\times 10^{-2}$ \\
 10 & -1.66838367$\times 10^{-4}$ & -4.59923079$\times 10^{-4}$ & & \\
 12 & -7.78978452$\times 10^{-5}$ & -2.59993308$\times 10^{-4}$ & & \\
\end{tabular}
\end{ruledtabular}
\end{table*}

\begin{table*}
\squeezetable
\caption{Series coefficients for two branches of magnon dispersion
$\omega^{\pm} (k_x, k_y) =$
$ J \sum_{k,n,m} a_{k,n,m} x^{k}
 [\cos (m k_x) \cos (n k_y) + \cos (n k_x) \cos (m k_y) ]/2  $ on the square lattice.
 Nonzero coefficients $a_{k,n,m}$
up to order $k=10$ are listed.}\label{tabdimgap}
\begin{ruledtabular}
\begin{tabular}{rl|rl|rl|rl}
\multicolumn{1}{c}{(k,n,m)} &\multicolumn{1}{c|}{$a_{k,n,m}$}
&\multicolumn{1}{c}{(k,n,m)} &\multicolumn{1}{c|}{$a_{k,n,m}$}
&\multicolumn{1}{c}{(k,n,m)} &\multicolumn{1}{c|}{$a_{k,n,m}$}
&\multicolumn{1}{c}{(k,n,m)} &\multicolumn{1}{c}{$a_{k,n,m}$} \\
\hline
\multicolumn{8}{c}{$\omega^-/J$} \\
 ( 0, 0, 0) & ~2.000000000    &( 8, 0, 2) & ~3.713762427$\times 10^{-3}$ &( 6, 3, 3) & -8.349006559$\times 10^{-3}$ &( 8, 3, 5) & -2.354734575$\times 10^{-3}$  \\
 ( 2, 0, 0) & -4.000000000$\times 10^{-1}$ &(10, 0, 2) & ~1.371857188$\times 10^{-3}$ &( 8, 3, 3) & -2.461673445$\times 10^{-3}$ &(10, 3, 5) & -1.492768067$\times 10^{-3}$  \\
 ( 4, 0, 0) & ~6.325595238$\times 10^{-2}$ &( 4, 2, 2) & -5.812500000$\times 10^{-2}$ &(10, 3, 3) & -5.509512597$\times 10^{-4}$ &( 8, 2, 6) & -1.177367287$\times 10^{-3}$  \\
 ( 6, 0, 0) & -5.757990909$\times 10^{-3}$ &( 6, 2, 2) & -8.531165228$\times 10^{-3}$ &( 6, 2, 4) & -1.252350984$\times 10^{-2}$ &(10, 2, 6) & -1.124360790$\times 10^{-3}$ \\
 ( 8, 0, 0) & -1.507778153$\times 10^{-4}$ &( 8, 2, 2) & ~1.296551584$\times 10^{-3}$ &( 8, 2, 4) & -4.921900666$\times 10^{-3}$ &( 8, 1, 7) & -3.363906535$\times 10^{-4}$ \\
 (10, 0, 0) & ~4.027910703$\times 10^{-4}$ &(10, 2, 2) & -7.669154691$\times 10^{-5}$ &(10, 2, 4) & -1.312852923$\times 10^{-3}$ &(10, 1, 7) & -5.883082637$\times 10^{-4}$ \\
 ( 2, 1, 1) & -1.000000000    &( 4, 1, 3) & -7.750000000$\times 10^{-2}$ &( 6, 1, 5) & -5.009403935$\times 10^{-3}$ &( 8, 0, 8) & -2.102441585$\times 10^{-5}$ \\
 ( 4, 1, 1) & ~7.619047619$\times 10^{-2}$ &( 6, 1, 3) & -2.053764395$\times 10^{-2}$ &( 8, 1, 5) & -3.914975780$\times 10^{-3}$ &(10, 0, 8) & -9.061290707$\times 10^{-5}$ \\
 ( 6, 1, 1) & ~8.413191563$\times 10^{-3}$ &( 8, 1, 3) & ~1.132685760$\times 10^{-3}$ &(10, 1, 5) & -1.582142621$\times 10^{-3}$ &(10, 5, 5) & -3.071949277$\times 10^{-4}$ \\
 ( 8, 1, 1) & ~3.139563846$\times 10^{-3}$ &(10, 1, 3) & -6.338716677$\times 10^{-4}$ &( 6, 0, 6) & -4.174503279$\times 10^{-4}$ &(10, 4, 6) & -5.119915461$\times 10^{-4}$ \\
 (10, 1, 1) & ~2.939474184$\times 10^{-3}$ &( 4, 0, 4) & -9.687500000$\times 10^{-3}$ &( 8, 0, 6) & -8.955696061$\times 10^{-4}$ &(10, 3, 7) & -2.925665978$\times 10^{-4}$ \\
 ( 2, 0, 2) & -5.000000000$\times 10^{-1}$ &( 6, 0, 4) & -8.507941330$\times 10^{-3}$ &(10, 0, 6) & -6.494183735$\times 10^{-4}$ &(10, 2, 8) & -1.097124742$\times 10^{-4}$ \\
 ( 4, 0, 2) & -5.000000000$\times 10^{-3}$ &( 8, 0, 4) & -1.541337062$\times 10^{-3}$ &( 8, 4, 4) & -1.471709109$\times 10^{-3}$ &(10, 1, 9) & -2.438054981$\times 10^{-5}$ \\
 ( 6, 0, 2) & -3.672253722$\times 10^{-3}$ &(10, 0, 4) & -6.373243437$\times 10^{-4}$ &(10, 4, 4) & -7.998702769$\times 10^{-4}$ &(10, 0,10) & -1.219027491$\times 10^{-6}$ \\
\hline
\multicolumn{8}{c}{$\omega^+/J$} \\
 ( 0, 0, 0) & ~4.000000000    &( 8, 0, 2) & -3.881904848$\times 10^{-4}$ &( 6, 3, 3) & -8.349006559$\times 10^{-3}$ &( 8, 3, 5) & -2.354734575$\times 10^{-3}$ \\
 ( 2, 0, 0) & ~1.000000000$\times 10^{-1}$ &(10, 0, 2) & ~2.303862243$\times 10^{-3}$ &( 8, 3, 3) & -2.422897285$\times 10^{-3}$ &(10, 3, 5) & -1.498879257$\times 10^{-3}$ \\
 ( 4, 0, 0) & ~1.112767857$\times 10^{-1}$ &( 4, 2, 2) & -5.812500000$\times 10^{-2}$ &(10, 3, 3) & -4.416530724$\times 10^{-4}$ &( 8, 2, 6) & -1.177367287$\times 10^{-3}$ \\
 ( 6, 0, 0) & ~1.200877715$\times 10^{-2}$ &( 6, 2, 2) & -8.773526841$\times 10^{-3}$ &( 6, 2, 4) & -1.252350984$\times 10^{-2}$ &(10, 2, 6) & -1.124356167$\times 10^{-3}$ \\
 ( 8, 0, 0) & ~2.554363932$\times 10^{-3}$ &( 8, 2, 2) & ~3.223616293$\times 10^{-4}$ &( 8, 2, 4) & -4.899434792$\times 10^{-3}$ &( 8, 1, 7) & -3.363906535$\times 10^{-4}$ \\
 (10, 0, 0) & ~2.741930556$\times 10^{-3}$ &(10, 2, 2) & ~9.155008757$\times 10^{-4}$ &(10, 2, 4) & -1.084169221$\times 10^{-3}$ &(10, 1, 7) & -5.868496963$\times 10^{-4}$ \\
 ( 2, 1, 1) & -1.000000000    &( 4, 1, 3) & -7.750000000$\times 10^{-2}$ &( 6, 1, 5) & -5.009403935$\times 10^{-3}$ &( 8, 0, 8) & -2.102441585$\times 10^{-5}$ \\
 ( 4, 1, 1) & ~8.333333333$\times 10^{-2}$ &( 6, 1, 3) & -2.005399403$\times 10^{-2}$ &( 8, 1, 5) & -3.948827468$\times 10^{-3}$ &(10, 0, 8) & -9.039423754$\times 10^{-5}$ \\
 ( 6, 1, 1) & ~1.081211454$\times 10^{-2}$ &( 8, 1, 3) & -1.215032805$\times 10^{-3}$ &(10, 1, 5) & -1.415199959$\times 10^{-3}$ &(10, 5, 5) & -3.071949277$\times 10^{-4}$ \\
 ( 8, 1, 1) & -1.064573032$\times 10^{-3}$ &(10, 1, 3) & ~9.174112292$\times 10^{-4}$ &( 6, 0, 6) & -4.174503279$\times 10^{-4}$ &(10, 4, 6) & -5.119915461$\times 10^{-4}$ \\
 (10, 1, 1) & ~2.773744329$\times 10^{-3}$ &( 4, 0, 4) & -9.687500000$\times 10^{-3}$ &( 8, 0, 6) & -9.043403076$\times 10^{-4}$ &(10, 3, 7) & -2.925665978$\times 10^{-4}$ \\
 ( 2, 0, 2) & -5.000000000$\times 10^{-1}$ &( 6, 0, 4) & -8.144935564$\times 10^{-3}$ &(10, 0, 6) & -6.249105162$\times 10^{-4}$ &(10, 2, 8) & -1.097124742$\times 10^{-4}$ \\
 ( 4, 0, 2) & -2.047619048$\times 10^{-2}$ &( 8, 0, 4) & -2.188469498$\times 10^{-3}$ &( 8, 4, 4) & -1.471709109$\times 10^{-3}$ &(10, 1, 9) & -2.438054981$\times 10^{-5}$ \\
 ( 6, 0, 2) & ~9.929231541$\times 10^{-3}$ &(10, 0, 4) & -2.155536602$\times 10^{-4}$ &(10, 4, 4) & -8.049648609$\times 10^{-4}$ &(10, 0,10) & -1.219027491$\times 10^{-6}$ \\
\end{tabular}
\end{ruledtabular}
\end{table*}

\begin{table*}
\squeezetable
\caption{Comparison of some numerical estimates obtained by different
methods for the ground-state energy $E_0/NJ$, the magnetization $M_1$, and the
energy gap $\omega^+({\bf k}=0)$, and  $\alpha$ at $x=1$.
}\label{tab_res}
\begin{ruledtabular}
\begin{tabular}{|rllll|} 
\multicolumn{1}{|c}{method} & \multicolumn{1}{c}{$E_0/NJ$} & \multicolumn{1}{c}{$M_1$}
 & \multicolumn{1}{c}{$\omega^+({\bf k}=0)$} & \multicolumn{1}{c|}{$\alpha$}  \\
\hline
\multicolumn{5}{|c|}{linear chain}\\
series (present work) &  -0.7270467(10) &  0.292487(6)  &   1.7591(6) &  1.66(3) \\ 
DMRG\cite{pat97}      &  -0.72704       &  0.29248      &   &   \\
QMC\cite{yam98}       &                 &               &    1.75914(1) &  1.48(4) \\
1st order SWT         &    -0.718228    &   0.195113   &     1   &        2     \\
2nd order SWT         &    -0.730420    &   0.316344   & 1.67556 &    1.5218	\\
3rd order SWT\cite{iva98}& -0.727161    &   0.293884   &         &         	\\
\hline
\multicolumn{5}{|c|}{square lattice}\\
series (present work) &  -1.2065125(10)&   0.40206(1)   &   2.5775(8)  & 1.80(4)   \\
1st order SWT         &  -1.20027      &   0.390762     &   2        &   2      \\
2nd order SWT         &  -1.20731	   &   0.401293	  &	2.52798  &	 1.87255 \\
\hline
\multicolumn{5}{|c|}{simple cubic lattice}\\
series (present work) &  -1.693375(7)  &   0.44450(3)  &   3.505(10)  &   1.69(8)   \\
1st order SWT         &  -1.68903       &  0.441852    &   3    &  2   \\
2st order SWT         &  -1.69371       &  0.444025    &   3.45069   &  1.95157   \\
\end{tabular}
\end{ruledtabular}
\end{table*}

\end{document}